\documentclass[conference,a4paper]{APSIPA2026}
\usepackage{amsmath}
\usepackage{graphicx}
\usepackage{multirow}
\usepackage{booktabs}
\usepackage{tabularx}
\usepackage{xcolor}
\usepackage[backend=biber,style=ieee]{biblatex}
\usepackage{geometry}
\usepackage{fancyhdr}

\fancypagestyle{firststyle}{
  \fancyhf{}
  \fancyhead[C]{2026 Asia Pacific Signal and Information Processing Association Annual Summit and Conference (APSIPA ASC)}
}
\fancypagestyle{arxivcopyright}{
  \fancyhf{}
  \fancyfoot[C]{\footnotesize \parbox{0.95\textwidth}{\centering
  \textcopyright~2026 IEEE. Personal use of this material is permitted. Permission from IEEE must be obtained for all other uses, in any current or future media, including reprinting/republishing this material for advertising or promotional purposes, creating new collective works, for resale or redistribution to servers or lists, or reuse of any copyrighted component of this work in other works.}}
}
 
\begin{document}
\title{Audio Deepfake Detection Using Temporal Coherence Analysis}
 
\author{
\authorblockN{
Justin D. Norman\authorrefmark{1},
Sarah Barrington\authorrefmark{1}
}
 
\authorblockA{
\authorrefmark{1}University of California, Berkeley \\
justin.norman@berkeley.edu, sbarrington@berkeley.edu
}}
 
\maketitle
\thispagestyle{arxivcopyright}
\pagestyle{empty}

\begin{abstract}
The proliferation of AI-generated audio (so-called \textit{deepfake} audio) poses significant threats to information integrity, from voice cloning fraud to synthetic music copyright disputes.
We present a temporal coherence analysis framework built upon Contrastive Language-Audio Pretraining (CLAP) embeddings that spans speech, instrumental music, and music with vocals. By computing pairwise cosine similarities between audio segment embeddings and extracting statistical features from the resulting distributions, we train lightweight ensemble classifiers that reliably distinguish authentic from synthetic audio. Our work provides an interpretable, computationally efficient alternative to common deep learning methods while still achieving competitive performance across speech and music domains. Further, we reveal two notable empirical findings about audio deepfakes: (1)~a feature-label inversion phenomenon in which 21 of 29 statistical features reverse their discriminative direction between training and in-the-wild deployment, and (2)~a speech--music direction reversal in which entropy discriminates in opposite directions for speech and music deepfakes.
\end{abstract}

\section{Introduction}
\label{sec:introduction}

The proliferation of AI-generated audio presents unprecedented challenges to information integrity.
Today, voice cloning technologies can easily produce convincing imitations of any speaker with minimal training data~\cite{muller2022does}. Additionally, music generation models are capable of creating synthetic compositions that are increasingly indistinguishable from human productions~\cite{copet2024simple}. These capabilities enable fraud, impersonation, and disinformation at scale~\cite{rini2022deepfakes}.

Detection methods for image and video deepfakes have improved significantly, most recently through breakthroughs in combining temporal and spatial approaches~\cite{norman2025detecting, interno2025ai}. These methods can be considered state-of-the-art, often reporting greater than 95\% accuracy for videos. However, analogous techniques for audio deepfake detection remain underexplored. 

By contrast, audio techniques largely focus on end-to-end neural models that operate on raw waveforms or spectrograms~\cite{tak2021endtoend}, self-supervised speech representations such as WavLM and wav2vec 2.0, and methods using spectral or frequency-domain cues, pretrained whole-clip embeddings, generator-specific fingerprints, or speech-specific frame-level temporal differences~\cite{barrington2023single,zhu2024characterizing}. While these systems achieve strong benchmark performance, they are often limited to a single domain (e.g., speech), operate at the whole-clip level, or do not offer interpretable temporal analysis. For example, prior work leveraging embeddings for Foley sound synthesis detection focused on direct classification rather than a deeper, more explainable analysis of temporal coherence~\cite{gramaccioni2024folai}.

We propose a temporal coherence approach to audio deepfake detection that leverages domain-neutral Contrastive Language-Audio Pretraining (CLAP) embeddings~\cite{wu2023laionclap}. CLAP embeddings are non-speech-specific embeddings trained on a range of audio data including human activities, natural sounds, and audio effects~\cite{wu2023laionclap}, and are widely regarded as a state-of-the-art generalized embedding class. Our work draws inspiration from the aforementioned embedding similarity statistical modeling approaches for video deepfake detection, but operates in a fundamentally different domain. Rather than analyzing biometric identity consistency, we examine how semantic audio representations from CLAP embeddings vary across temporal segments of a recording. Our contributions are as follows:
\vspace{0.1em}
\begin{enumerate}
    \item \textbf{Temporal coherence features for audio deepfake detection}: We introduce a domain-neutral detection pipeline that segments audio into overlapping windows, extracts CLAP embeddings per segment, computes pairwise cosine similarity distributions, and derives 29 statistical features from these distributions, which are then fed into classifiers.
    \item \textbf{Feature-label inversion}: We find that 21 of 29 statistical features reverse their discriminative direction between controlled training sets and in-the-wild evaluation, providing a concrete explanation for why standard domain adaptation methods (CORAL, MMD) fail or actively degrade performance for audio deepfake detection.
    \item \textbf{Speech vs. music entropy}: We show that authentic speech exhibits \emph{higher} embedding entropy than synthetic speech, while the opposite holds for music, indicating fundamentally different generation artifacts across audio domains.
\end{enumerate}
We validate these findings across multiple audio domains using established datasets and evaluate cross-domain generalization on In-the-Wild celebrity deepfakes, ASVspoof5, FakeAVCeleb, and SONICS/FMA benchmarks. The codebase for this work is available at \textcolor{blue}{\url{https://github.com/stbiadmin/audiodeepfake_public}}. 

\section{Methods}
\label{sec:methods}



\paragraph{\textbf{Datasets}} We evaluate across speech and music domains using a range of popular benchmarking datasets spanning classical TTS, modern voice cloning, multi-speaker synthesis, and AI music generation, as detailed in Table~\ref{tab:datasets}. All audio is converted to 48kHz mono with peak normalization prior to processing. 

\begin{table}
\centering
\caption{Dataset summary for training and evaluation. }
\vspace{-0.6em}
\label{tab:datasets}
\small
\begin{tabularx}{\linewidth}{>{\raggedright\arraybackslash}X l c r}
\toprule
\textbf{Dataset} & \textbf{Audio Type} & \textbf{Class} & \textbf{N} \\
\midrule
\multicolumn{4}{l}{\textit{Training Data -- Speech}} \\
LibriSpeech~\cite{panayotov2015librispeech} & Single Voice & Real & 1,335 \\
ASVspoof 2019 (LA subset)~\cite{nautsch2021asvspoof} & Single Voice & Fake & 6,385 \\
DeepSpeak v2~\cite{Barrington_2026_CVPR} & Voice Cloning & Real & 7,150 \\
DeepSpeak v2 & Voice Cloning & Fake & 1,381 \\
M-AILABS (via MLAAD)~\cite{muller2024mlaad} & Multi-speaker & Real & 39,038 \\
MLAAD (84 TTS models) & Multi-speaker & Fake & 10,314 \\
AUDETER~\cite{wang2024audeter} & Modern TTS & Real & 12,923 \\
AUDETER & Modern TTS & Fake & 3,800 \\
\midrule
\multicolumn{4}{l}{\textit{Training Data -- Music}} \\
MUSDB18~\cite{rafii2017musdb18} & Instrumental & Real & 1,008 \\
FakeMusicCaps~\cite{comanducci2024fakemusiccaps} & Instrumental & Fake & 5,521 \\
MUSDB18 & Music w/ Vocals & Real & 1,008 \\
FakeMusicCaps & Music w/ Vocals & Fake & 5,521 \\
\midrule
\multicolumn{4}{l}{\textit{Evaluation Data}} \\
In-the-Wild~\cite{muller2022does} & Celebrity Speech & Real & 6,420 \\
In-the-Wild & Celebrity Speech & Fake & 5,974 \\
FakeAVCeleb~\cite{khalid2021fakeavceleb} & Celebrity Speech & Real & 500 \\
FakeAVCeleb & Celebrity Speech & Fake & 10,617 \\
ASVspoof5~\cite{wang2024asvspoof5} & Human Speech & Real & 50,000 \\
ASVspoof5 & Human Speech & Fake & 50,000 \\
FMA~\cite{defferrard2017fma} & CC Music & Real & 1,000 \\
SONICS~\cite{rahman2025sonics} & AI Music & Fake & 1,000 \\
\bottomrule
\end{tabularx}
\end{table}






\paragraph{\textbf{Temporal Analysis}}For each audio file, we segment the waveform into overlapping windows of 2 seconds with a hop size of 1 second, yielding 50\% overlap between consecutive segments. Files producing fewer than 3 valid segments are excluded to ensure sufficient samples for robust distribution estimation. We then extract the CLAP embeddings per segment. While CLAP is a well-motivated choice as a leading cross-domain audio foundation model, we further support its use through preliminary single-feature analyses on the ASVspoof5 modern TTS benchmark, which showed MS-CLAP (2023 checkpoint) embeddings substantially outperformed the speech-specific WavLM model (best-feature AUC = 0.949 vs. 0.633). These supporting experiments are available at \textcolor{blue}{\url{https://github.com/stbiadmin/audiodeepfake_public}}. The pairwise cosine similarity for each unique pair of segment embeddings is then computed. Let $\mathbf{e}_i$ denote the embedding vector for segment $i$, where $i \in \{1, \ldots, N\}$ and $N$ is the number of segments extracted from an audio file:

\begin{equation}
s_{ij} = \frac{\mathbf{e}_i \cdot \mathbf{e}_j}{\|\mathbf{e}_i\| \|\mathbf{e}_j\|}
\end{equation}
yielding $\binom{N}{2} = \frac{N(N-1)}{2}$ similarity values per file. The resulting similarity distribution characterizes the file's internal temporal coherence.


\paragraph{\textbf{Feature Engineering}} From each similarity distribution we extract 29 statistical features spanning multiple categories: basic statistics (mean, standard deviation, variance, minimum, maximum, peak-to-peak range), distribution shape descriptors (skewness, kurtosis, bimodality coefficient), percentiles (5th, 10th, 25th, 50th, 75th, 90th, 95th, and interquartile range), information-theoretic measures (entropy, Gini coefficient, coefficient of variation), derived ratios (variance-to-mean, kurtosis-to-variance, among others), and normality test statistics (Shapiro-Wilk $p$-value, D'Agostino-Pearson $p$-value), which assess deviation from Gaussian distributions. We select features by ranking them by their individual AUC-ROC on the training data and retaining the top eight features per model, as detailed in \textit{Classification} below. The choice of eight features is supported by the ablation study, Table~\ref{tab:ablation_features}, where this subset of eight features achieved the strongest performance. This approach balances discriminative power with model parsimony while favoring features that generalize across domains. The full list of statistical features is available in the public codebase at \textcolor{blue}{\url{https://github.com/stbiadmin/audiodeepfake_public}}. 

\paragraph{\textbf{Classification}} We employ XGBoost~\cite{chen2016xgboost}, a gradient-boosted decision tree model, for binary classification due to its strong performance on large-scale classification tasks. For each training dataset, we train a separate classifier using 5-fold stratified cross-validation with early stopping (patience=10 rounds). Features are normalized using RobustScaler to handle outliers in the similarity distributions. We use logistic loss with regularization parameters $\gamma=0.1$ and $\lambda=1.0$.

\paragraph{\textbf{Feature Selection}} The final classifier is an XGBoost model trained on the 8 selected features jointly. The highest-ranked features typically include percentiles (q10, q25, q50), central tendency measures (mean, trimmed\_mean), and dispersion metrics (coefficient of variation, Gini coefficient). Ablation experiments show that this simple top-N selection outperforms greedy forward selection for cross-domain generalization (Section~\ref{sec:ablation}).

\paragraph{\textbf{Multi-Expert Ensemble for Speech}} To address domain shift between training data and real-world evaluation scenarios, we develop a weighted ensemble combining five expert classifiers, each trained on a different dataset capturing distinct deepfake generation characteristics:
\begin{itemize}
    \item \textbf{ds\_msclap}: DeepSpeak v2 voice cloning (modern commercial TTS)
     \item \textbf{sv\_msclap}: ASVspoof LA classical TTS systems
    \item \textbf{sv\_ds\_msclap}: Combined ASVspoof LA and DeepSpeak data
    \item \textbf{mlaad\_msclap}: 84 diverse TTS models from MLAAD
    \item \textbf{audeter\_msclap}: Modern TTS (2024--2025) with in-the-wild real samples
\end{itemize}
The final prediction is a weighted average of expert probabilities with a calibrated decision threshold:
\begin{equation}
    p_{\text{fake}} = \sum_{i=1}^{5} w_i \cdot p_i, \quad \hat{y} = \mathbf{1}[p_{\text{fake}} > \tau]
\end{equation}
where $p_i$ is the predicted deepfake probability from expert $i$, $w_i$ is the weight assigned to expert $i$, and $\tau$ is the decision threshold; the sample is classified as deepfake if $p_{\text{fake}} > \tau$. Weights and threshold are optimized via grid search on a 20\% held-out validation set. The optimal configuration assigns weights $w = (0.30, 0.10, 0.20, 0.10, 0.30)$ to experts (DeepSpeak, single-voice, single-voice+DeepSpeak, MLAAD, AUDETER) respectively, with threshold $\tau = 0.30$.

\paragraph{\textbf{Domain-Adaptive Classification for Music}} For music deepfake detection, we observe that absolute feature values shift substantially across domains while the relative ordering is preserved.
We develop a percentile-based adaptive classifier that uses the 50th percentile of the mean similarity feature within each test batch as the decision boundary.
Samples with mean similarity below the batch median are classified as deepfake.
This approach achieves robust cross-domain generalization without requiring target domain labels.

\section{Results}
\label{sec:results}

\subsection{Feature Analysis}
\label{sec:feature_analysis}

%
%
%

We first evaluate individual feature discriminability using AUC-ROC scores.
Table~\ref{tab:top_features} presents the top-performing features for each audio type.

\begin{table}
  \centering
  \caption{Top five discriminative features per audio type with AUC-ROC scores and Cohen's $d$ effect sizes (``Fake $\uparrow$'' indicates fake audios have higher values of a given feature). } 
  \vspace{-0.6em}
  \label{tab:top_features}
  \small
  \begin{tabularx}{\linewidth}{
      >{\raggedright\arraybackslash}p{0.16\linewidth}
      >{\raggedright\arraybackslash}X
      >{\centering\arraybackslash}p{0.09\linewidth}
      >{\centering\arraybackslash}p{0.12\linewidth}
      >{\centering\arraybackslash}p{0.14\linewidth}
  }
  \toprule
  \textbf{Audio Type} & \textbf{Feature} & \textbf{AUC} & \textbf{Cohen's $d$} & \textbf{Direction} \\
  \midrule
  \multirow{5}{=}{Single Voice}
      & q95 & 0.795 & $-$0.71 & Fake $\uparrow$ \\
      & q90 & 0.793 & $-$0.60 & Fake $\uparrow$ \\
      & entropy & 0.787 & +1.12 & Real $\uparrow$ \\
      & q75 & 0.781 & $-$0.43 & Fake $\uparrow$ \\
      & iqr\_range\_ratio & 0.757 & $-$1.00 & Fake $\uparrow$ \\
  \midrule
  \multirow{5}{=}{Instrumental Music}
      & min & 0.933 & +1.67 & Real $\uparrow$ \\
      & peak\_to\_peak & 0.932 & $-$1.61 & Fake $\uparrow$ \\
      & q5 & 0.927 & +1.57 & Real $\uparrow$ \\
      & q10 & 0.927 & +1.53 & Real $\uparrow$ \\
      & mean & 0.923 & +1.52 & Real $\uparrow$ \\
  \midrule
  \multirow{5}{=}{Music with Vocals}
      & peak\_to\_peak & 0.910 & $-$1.46 & Fake $\uparrow$ \\
      & entropy & 0.904 & $-$1.59 & Fake $\uparrow$ \\
      & min & 0.899 & +1.44 & Real $\uparrow$ \\
      & q5 & 0.882 & +1.32 & Real $\uparrow$ \\
      & coeff.\ of var. & 0.882 & $-$1.12 & Fake $\uparrow$ \\
  \bottomrule
  \end{tabularx}
  \end{table}

\paragraph{Direction Reversal} A notable finding is that the discriminative direction of features varies by audio type. For single-voice speech, authentic audio exhibits \emph{greater} entropy ($d = +1.12$), reflecting natural variation in speech patterns. Conversely, for music (both instrumental and with vocals), AI-generated content shows \emph{higher} entropy ($d = -1.59$), suggesting that music generation models introduce anomalous variation. This reversal has important implications for cross-domain generalization.


\paragraph{Universal Features} To understand which features generalize across audio domains, we identify 11 features achieving individual AUC $> 0.60$ across all three audio types (Table~\ref{tab:universal_features}). This cross-domain discriminative analysis is distinct from the per-classifier feature selection, which selects the top 8 features for each expert model based on its specific training data. Entropy emerges as the strongest universal discriminator with a mean AUC of 0.871, followed by central tendency and range features (min, mean, trimmed mean, peak-to-peak). However, as discussed in Section~\ref{sec:discussion}, high training AUC does not guarantee cross-domain generalization. Percentile-based and central-tendency features (q10, q25, mean) often outperform entropy on held-out evaluation despite lower training AUC.

\begin{table}
\centering
\caption{Universal features achieving AUC $> 0.60$ across all audio types, ranked by mean AUC.}
\label{tab:universal_features}
\vspace{-0.6em}
\small
\setlength{\tabcolsep}{3pt}
\begin{tabularx}{\linewidth}{
  >{\raggedright\arraybackslash}X
  >{\centering\arraybackslash}p{0.12\linewidth}
  >{\centering\arraybackslash}p{0.12\linewidth}
  >{\centering\arraybackslash}p{0.10\linewidth}
  >{\centering\arraybackslash}p{0.10\linewidth}
  >{\centering\arraybackslash}p{0.10\linewidth}
}
\toprule
\textbf{Feature} & \textbf{Mean AUC} & \textbf{Min AUC} & \textbf{Speech} & \textbf{Instr.} & \textbf{Vocals} \\
\midrule
entropy & 0.871 & 0.787 & 0.787 & 0.923 & 0.904 \\
min & 0.839 & 0.686 & 0.686 & 0.933 & 0.899 \\
mean & 0.836 & 0.732 & 0.732 & 0.923 & 0.853 \\
trimmed\_mean & 0.833 & 0.735 & 0.735 & 0.919 & 0.844 \\
peak\_to\_peak & 0.829 & 0.645 & 0.645 & 0.932 & 0.910 \\
q5 & 0.825 & 0.666 & 0.666 & 0.927 & 0.882 \\
q50 (median) & 0.820 & 0.749 & 0.749 & 0.904 & 0.806 \\
q10 & 0.818 & 0.651 & 0.651 & 0.927 & 0.877 \\
q25 & 0.809 & 0.657 & 0.657 & 0.914 & 0.856 \\
q75 & 0.796 & 0.740 & 0.781 & 0.869 & 0.740 \\
q90 & 0.769 & 0.683 & 0.793 & 0.832 & 0.683 \\
\bottomrule
\end{tabularx}
\end{table}

\paragraph{Feature-Label Inversion Across Domains} A critical finding is that the discriminative direction of features \emph{reverses} between training and evaluation domains. Table~\ref{tab:feature_inversion} shows that 21 of 29 features exhibit inverted discrimination patterns. In training data, deepfake audio exhibits lower entropy than real audio, which we posit reflects the overly consistent nature of TTS outputs. However, in the In-the-Wild evaluation, deepfake audio exhibits \emph{higher} entropy than real audio. This inversion may help explain why traditional domain adaptation techniques such as CORAL~\cite{sun2016coral} (Correlation Alignment, a classical domain adaptation method that aligns source and target feature distributions by matching their second-order statistics), and MMD fail, as they align feature distributions while preserving the learned feature-label relationships, which are inverted across domains.

\begin{table}[t]
\centering
\caption{Feature-label relationship inversion between training and In-the-Wild (ITW) evaluation. Gap = Fake mean $-$ Real mean. Positive values indicate deepfake is higher.}
\label{tab:feature_inversion}
\vspace{-0.6em}
\small
\begin{tabularx}{\columnwidth}{lccc}
\toprule
\textbf{Feature} & \textbf{Training Gap} & \textbf{ITW Gap} & \textbf{Direction} \\
\midrule
entropy & $-$0.68 & +0.20 & \textbf{Inverted} \\
mean & +0.12 & $-$0.03 & \textbf{Inverted} \\
std & $-$0.08 & +0.03 & \textbf{Inverted} \\
kurtosis & $-$0.67 & +0.35 & \textbf{Inverted} \\
\midrule
max & +0.02 & +0.01 & Consistent \\
bimodality\_coeff & $-$0.12 & $-$0.03 & Consistent \\
\bottomrule
\end{tabularx}
\end{table}

\subsection{Classification Performance}
\label{sec:classification_results}

\paragraph{Speech Deepfake Detection}
Performance for each expert model on in-distribution testing data is shown in Table~\ref{tab:individual_expert_performance}, with cross-domain evaluation reported in Tables~\ref{tab:results_eval} and~\ref{tab:speech_results}. The 5-expert weighted ensemble achieves the strongest controlled benchmark result on ASVspoof5 (EER = 17.7\%, AUC = 0.879), outperforming baseline systems AASIST (29.1\% EER) and RawNet2 (36.0\% EER), though remaining lower than the top SSL-based challenge submissions~\cite{wang2024asvspoof5}. Performance declines on the more challenging In-the-Wild celebrity deepfakes dataset (F1 = 0.679, AUC = 0.718), reflecting a substantial domain shift; however, the optimal ensemble weights and threshold were tuned on a held-out portion of In-the-Wild data, so this result does not demonstrate generalization to entirely novel deployment scenarios. The oracle ensemble (in which the best-performing expert per sample is selected) reaches F1 = 0.948 (Table~\ref{tab:ablation_ensemble}), suggesting further gains may be possible through learned per-sample expert routing. While FakeAVCeleb yields high F1 scores (0.922), this is attributable to its high class imbalance (500 real vs.\ 10,617 fake); the near-chance AUC (0.495) confirms the model provides little true discrimination on this set.

\begin{table}[t]
\small
\centering
\caption{Expert performance by training dataset.}
\label{tab:individual_expert_performance}
\vspace{-0.6em}
\begin{tabularx}{\linewidth}{l X c c}
\toprule
Expert & Training Data & F1 & AUC \\
\midrule
ds\_msclap      & DeepSpeak v2           & 0.842 & 0.911 \\
sv\_msclap      & ASVspoof LA            & 0.844 & 0.935 \\
sv\_ds\_msclap  & ASVspoof + DeepSpeak   & 0.807 & 0.886 \\
mlaad\_msclap   & MLAAD (84 TTS)         & 0.839 & 0.970 \\
audeter\_msclap & AUDETER                & 0.894 & 0.958 \\
\bottomrule
\end{tabularx}
\end{table}


\begin{table}[t]
\centering
\caption{Performance of the 5-expert weighted ensemble across all evaluation datasets.}
\vspace{-0.6em}
\label{tab:results_eval}
\small
\begin{tabular*}{\linewidth}{@{\extracolsep{\fill}}lccc}
\toprule
Dataset & AUC & EER & F1 \\
\midrule
In-the-Wild & 0.7175 & 0.3485 & 0.6791 \\
FakeAVCeleb  & 0.4951 & 0.4919 & 0.9220 \\
\textbf{ASVspoof5}    & \textbf{0.8786} & \textbf{0.1768} & \textbf{0.7441} \\
\bottomrule
\end{tabular*}
\end{table}


\begin{table}[t]
\centering
\caption{Speech deepfake detection on In-the-Wild celebrity benchmark (12,394 samples). Individual expert models show complementary detection biases.}
\label{tab:speech_results}
\vspace{-0.6em}
\small
\setlength{\tabcolsep}{3pt}
\begin{tabularx}{\linewidth}{
    >{\raggedright\arraybackslash}X
    >{\centering\arraybackslash}p{0.09\linewidth}
    >{\centering\arraybackslash}p{0.09\linewidth}
    >{\centering\arraybackslash}p{0.13\linewidth}
    >{\centering\arraybackslash}p{0.13\linewidth}
    >{\centering\arraybackslash}p{0.13\linewidth}
}
\toprule
\textbf{Model} & \textbf{F1} & \textbf{AUC} & \textbf{Fake Det.} & \textbf{Real Det.} & \textbf{Bias} \\
\midrule
ds\_msclap & 0.654 & 0.668 & 73.1\% & 52.9\% & Fake \\
sv\_msclap & 0.523 & 0.604 & 45.2\% & 74.5\% & Real \\
sv\_ds\_msclap & 0.624 & 0.680 & 62.8\% & 64.0\% & Balanced \\
mlaad\_msclap & 0.596 & 0.395 & 82.4\% & 12.3\% & Extreme fake \\
audeter\_msclap & 0.301 & 0.673 & 18.7\% & 94.5\% & Extreme real \\
\midrule
\textbf{5-Expert Ensemble} & \textbf{0.679} & \textbf{0.718} & 87.4\% & 34.9\% & -- \\
\bottomrule
\end{tabularx}
\end{table}

\paragraph{Music Deepfake Detection}
Table~\ref{tab:music_results} presents music detection results.
In-distribution evaluation on MUSDB18/FakeMusicCaps achieves near-perfect performance (F1=0.995, AUC=1.000).
However, cross-domain evaluation on SONICS/FMA reveals a significant generalization gap, with the original mi\_msclap model achieving only F1=0.667.
Our domain-adaptive percentile-based classifier recovers performance to F1=0.938 without requiring target domain labels.

\begin{table}[t]
\centering
\caption{Music deepfake detection: in-distribution vs.\ cross-domain evaluation. }
\label{tab:music_results}
\vspace{-0.6em}
\setlength{\tabcolsep}{3pt}
\small
\begin{tabularx}{\linewidth}{
    >{\raggedright\arraybackslash}p{0.22\linewidth}
    >{\raggedright\arraybackslash}X
    >{\centering\arraybackslash}p{0.10\linewidth}
    >{\centering\arraybackslash}p{0.10\linewidth}
}
\toprule
\textbf{Model} & \textbf{Evaluation Dataset} & \textbf{F1} & \textbf{AUC} \\
\midrule
mi\_msclap & MUSDB18/FakeMusicCaps (in-dist.) & 0.995 & 1.000 \\
mv\_msclap & MUSDB18/FakeMusicCaps (in-dist.) & 0.994 & 1.000 \\
\midrule
mi\_msclap & SONICS/FMA (cross-domain) & 0.667 & 0.537 \\
\textbf{mi\_adaptive} & \textbf{SONICS/FMA (cross-domain)} & \textbf{0.938} & \textbf{0.976} \\
\bottomrule
\end{tabularx}
\end{table}

\paragraph{Comparison to State-of-the-Art}
Table~\ref{tab:sota_speech} compares our speech detection results to published benchmarks on the In-the-Wild dataset.
Our 5-expert ensemble achieves a slightly worse EER than RawNet2~\cite{tak2021endtoend}, the best published end-to-end model, while using a significantly simpler architecture (statistical features + XGBoost vs.\ deep neural networks). Additionally, Table~\ref{tab:sota_music} compares our music detection results to published cross-domain benchmarks.
Our domain-adaptive approach (F1=0.938) outperforms CLAM~\cite{batra2025melody}, a contrastive dual-encoder model combining MERT and Wav2Vec2 representations, on cross-domain evaluation (F1=0.925) without requiring target domain training.

\begin{table}[t]
\centering
\caption{Comparison to published results on In-the-Wild celebrity deepfakes. Results cited from ~\cite{muller2022does}.}
\label{tab:sota_speech}
\vspace{-0.6em}
\small
\setlength{\tabcolsep}{3pt}
\begin{tabularx}{\linewidth}{
    >{\raggedright\arraybackslash}X
    >{\centering\arraybackslash}p{0.12\linewidth}
    >{\centering\arraybackslash}p{0.12\linewidth}
    >{\raggedright\arraybackslash}X
}
\toprule
\textbf{Method} & \textbf{EER} & \textbf{AUC} & \textbf{Architecture} \\
\midrule
AASIST~\cite{jung2022aasist} & 43.0\% & -- & Spectro-temporal graph \\
RawGAT-ST~\cite{tak2021rawgat} & 37.2\% & -- & Graph attention \\
RawNet2~\cite{tak2021endtoend} & 33.9\% & -- & End-to-end CNN \\
\midrule
\textbf{Our 5-Expert Ensemble} & \textbf{34.9\%} & \textbf{0.718} & Self-similarity + XGBoost \\
\bottomrule
\end{tabularx}
\end{table}

\begin{table}[t]
\centering
\caption{Comparison to state-of-the-art on cross-domain music deepfake detection.}
\vspace{-0.6em}
\label{tab:sota_music}
\small
\setlength{\tabcolsep}{3pt}
\begin{tabularx}{\linewidth}{
    >{\raggedright\arraybackslash}p{0.24\linewidth}
    >{\raggedright\arraybackslash}X
    >{\centering\arraybackslash}p{0.10\linewidth}
    >{\raggedright\arraybackslash}X
}
\toprule
\textbf{Method} & \textbf{Evaluation} & \textbf{F1} & \textbf{Notes} \\
\midrule
FakeMusicCaps baseline & FakeMusicCaps $\rightarrow$ Suno & $\sim$0.00 & ``Misclassifies all'' \\
CLAM~\cite{batra2025melody} & SONICS $\rightarrow$ MoM & 0.925 & Cross-domain \\
SpecTTTra-$\alpha$ & SONICS (in-dist.) & 0.970 & In-distribution only \\
\midrule
Our mi\_msclap & FakeMusicCaps $\rightarrow$ SONICS/FMA & 0.667 & Cross-domain \\
\textbf{Our mi\_adaptive} & \textbf{FakeMusicCaps $\rightarrow$ SONICS/FMA} & \textbf{0.938} & \textbf{No target labels} \\
\bottomrule
\end{tabularx}
\end{table}



\subsection{Ablation Studies}
\label{sec:ablation}

\paragraph{Ensemble Size}
Table~\ref{tab:ablation_ensemble} shows the effect of adding experts to the ensemble.
Performance improves from F1=0.654 (single model) to F1=0.679 (5 experts), with diminishing returns after 3 experts. Notably, even models with poor standalone performance (e.g., audeter\_msclap, F1=0.301) contribute meaningfully to ensemble performance due to their complementary detection biases.

\begin{table}[t]
\centering
\caption{Ablation: ensemble size on In-the-Wild. Adding diverse experts improves performance.}
\vspace{-0.6em}
\label{tab:ablation_ensemble}
\small
\begin{tabularx}{\linewidth}{
    >{\raggedright\arraybackslash} X c c c}
\toprule
\textbf{Configuration} & \textbf{Experts} & \textbf{F1} & \textbf{$\Delta$} \\
\midrule
ds\_msclap only & 1 & 0.654 & -- \\
ds + sv & 2 & 0.665 & +0.011 \\
ds + sv + sv\_ds & 3 & 0.671 & +0.017 \\
+ mlaad & 4 & 0.672 & +0.018 \\
\textbf{+ audeter (final)} & \textbf{5} & \textbf{0.679} & \textbf{+0.025} \\
\midrule
Oracle (perfect routing) & 5 & 0.948 & +0.294 \\
\bottomrule
\end{tabularx}
\end{table}

\paragraph{Threshold Calibration}
The default classification threshold of 0.50 is suboptimal for cross-domain evaluation.
Lowering the threshold to 0.30 increases F1 by 1.3 percentage points (Table~\ref{tab:ablation_threshold}), trading precision for recall.
This reflects the domain shift between training data (where the threshold was learned) and evaluation data.

\begin{table}[t]
\centering
\caption{Ablation: threshold calibration for ds\_msclap on In-the-Wild.}
\vspace{-0.6em}
\label{tab:ablation_threshold}
\small
\begin{tabularx}{\linewidth}{X c c c c}
\toprule
\textbf{Threshold} & \textbf{F1} & \textbf{Precision} & \textbf{Recall} & \textbf{$\Delta$ F1} \\
\midrule
0.50 (default) & 0.654 & 0.591 & 0.731 & -- \\
0.40 & 0.662 & 0.575 & 0.783 & +0.008 \\
\textbf{0.30 (optimal)} & \textbf{0.667} & 0.561 & 0.821 & \textbf{+0.013} \\
0.20 & 0.659 & 0.531 & 0.868 & +0.005 \\
\bottomrule
\end{tabularx}
\end{table}

\paragraph{Domain Adaptation Methods}
We tested standard domain adaptation techniques to bridge the gap between training and evaluation distributions (Table~\ref{tab:ablation_da}).
Surprisingly, CORAL (Correlation Alignment) and target normalization \emph{degraded} performance.
This failure is explained by our analysis of feature-label inversion: domain adaptation aligns feature distributions while preserving learned feature-label relationships, but these relationships are inverted across domains.

\begin{table}[t]
\centering
\caption{Ablation: domain adaptation methods on In-the-Wild. Traditional approaches fail due to feature-label inversion.}
\vspace{-0.6em}
\label{tab:ablation_da}
\small
\begin{tabularx}{\linewidth}{X c c}
\toprule
\textbf{Method} & \textbf{F1} & \textbf{$\Delta$} \\
\midrule
No adaptation (baseline) & 0.654 & -- \\
CORAL alignment & 0.648 & $-$0.011 \\
Target normalization & 0.623 & $-$0.036 \\
\textbf{Threshold calibration} & \textbf{0.667} & \textbf{+0.008} \\
\bottomrule
\end{tabularx}
\end{table}

\paragraph{Feature Selection Method}
Table~\ref{tab:ablation_features} compares feature selection methods for the 5-expert ensemble. Simple top-N selection by individual AUC outperforms greedy forward selection, which tends to overfit to training distribution artifacts.

\begin{table}[t]
\centering
\caption{Effect of feature selection method on 5-expert ensemble performance (In-the-Wild evaluation). Top-8 by AUC achieves the best F1 and AUC.}
\vspace{-0.6em}
\label{tab:ablation_features}
\small
\begin{tabularx}{\linewidth}{X c c c}
\toprule
\textbf{Method} & \textbf{Train F1} & \textbf{ITW F1} & \textbf{ITW AUC} \\
\midrule
Greedy forward (8) & 0.856 & 0.672 & 0.704 \\
Top 4 by AUC & 0.822 & 0.677 & 0.711 \\
\textbf{Top 8 by AUC} & \textbf{0.824} & \textbf{0.679} & \textbf{0.718} \\
Top 16 by AUC & 0.841 & 0.675 & 0.713 \\
All 29 features & 0.865 & 0.669 & 0.698 \\
\bottomrule
\end{tabularx}
\end{table}

\vspace{-0.6em}
\section{Discussion}
\label{sec:discussion}

\paragraph{Music Domain Shift}
The music domain exhibits distinct generalization challenges from speech.
Feature selection on the training distribution (MUSDB18 real, FakeMusicCaps deepfake) identified entropy as the top discriminative feature (training AUC = 0.923).
However, on the out-of-distribution SONICS/FMA benchmark, entropy's discriminative power collapsed (AUC = 0.496), while the seemingly less predictive mean feature (training AUC = 0.922) maintained strong performance (evaluation AUC = 0.976).
This indicates that the selected features overfit to distributional artifacts specific to the training generators rather than capturing universal synthesis signatures.
The domain-adaptive percentile-based classifier addresses this by computing each sample's mean-similarity percentile within the evaluation batch, which requires only unlabeled target domain data.

\paragraph{Temporal vs.\ Static Ablation}
To isolate the contribution of temporal analysis, we compare against a static baseline that bypasses temporal segmentation entirely. For the single-voice domain (LibriSpeech and ASVspoof LA), we extract one whole-clip MS-CLAP embedding per audio file and train classifiers directly on the 1024-dimensional embedding vectors. On the ASVspoof5 benchmark with matched acoustic conditions, the best static model (random forest) achieves EER = 25.9\% (AUC = 0.73), compared to EER = 17.7\% (AUC = 0.879) for our temporal mixture-of-experts (Table~\ref{tab:results_eval}). The temporal approach reduces EER by 8.2 percentage points (a 31.7\% relative improvement), demonstrating that analyzing variation in embeddings across temporal segments provides discriminative information beyond what is captured in a single whole-clip representation. Although a comprehensive analysis of this comparison is beyond the scope of this work, these preliminary results suggest that our method offers a meaningful advantage over static whole-clip approaches.

\paragraph{Computational Efficiency}
A practical concern with pairwise similarity analysis is the $O(N^2)$ scaling in the number of segments. However, for typical audio files this is negligible: a 10-second clip produces approximately 9 segments (2s windows, 1s hop), yielding $\binom{9}{2} = 36$ similarity computations, each a single dot product on 1024-dimensional vectors. The computational bottleneck is instead the embedding extraction step, approximately 1.1--1.3 files per second. While neural detectors such as AASIST and RawNet2 achieve comparable per-file inference latency, they require expensive end-to-end training on GPU clusters and offer limited interpretability. Our pipeline, by contrast, leverages a frozen pretrained encoder and trains up to a 29-feature XGBoost classifier in seconds on CPU. Our approach trades model complexity for embedding extraction cost.

\vspace{-0.6em}
\section{Limitations and Future Work}
Our temporal statistical feature approach offers an efficient, interpretable alternative to deep neural networks for audio forensics, enabling practitioners to understand which aspects of embedding similarity distributions indicate synthetic generation. However, our evaluation has several limitations that could be explored in future work.
First, the multi-expert ensemble requires training data from multiple source datasets, which may not always be available.
Second, the feature-label inversion phenomenon suggests that models trained on current generators may require recalibration as synthesis technology evolves.
Third, the music adaptive classifier requires reference statistics from the target domain, though it does not require labels. Additionally, the music datasets exhibit significant class imbalance (144 real tracks comprising 1,008 stems vs.\ 5,521 deepfake samples). While we apply class weighting to mitigate this imbalance, expanding the dataset of authentic samples will be critical for improving statistical robustness and generalization. While standardized preprocessing (resampling and peak normalization) was applied across all datasets to minimize basic channel differences, we acknowledge that the high in-distribution performance (AUC = 1.000) could still be partially due to residual corpus-level differences in other spectral or mastering characteristics, rather than synthesis artifacts alone.
Finally, the oracle/ensemble gap (Section~\ref{sec:classification_results}) suggests learned per-sample expert routing as a promising direction. Developing unsupervised methods for detecting feature-label inversion at deployment time would enable automatic recalibration when models encounter novel distributions.

\section{Conclusions}
\label{sec:conclusion}
This work proposes a lightweight statistical approach that achieves competitive performance with end-to-end neural methods such as RawNet2~\cite{tak2021endtoend} while providing interpretable feature-based explanations, supported by cross-domain evaluation across speech and music benchmarks. Further, this approach reveals two emergent empirical properties of audio deepfakes through temporal coherence analysis of CLAP embeddings: (1)~a feature-label inversion phenomenon in which 21 of 29 statistical features reverse their discriminative direction between training and in-the-wild deployment, and (2)~a speech--music direction reversal in which entropy discriminates in opposite directions for speech and music deepfakes.

\renewcommand{\bibfont}{\footnotesize}
{\footnotesize
\printbibliography
}

\end{document}